\documentclass[prl,aps,preprint,amsmath]{revtex4}
\usepackage{graphicx,amsmath,units,xspace,amssymb}
\usepackage{graphicx}
\usepackage{dcolumn}
\usepackage{bm}
\usepackage{xcolor}
\newcommand{\RNum}[1]{\uppercase\expandafter{\romannumeral #1\relax}}
\newcolumntype{+}{!{\vrule width 2pt}}

\newlength\savedwidth

\begin{document}

\title{Achieving adjustable elasticity with non-affine to affine transition}






\author{Xiangying Shen$^{1,2,3\dagger}$, Chenchao Fang$^{1,3\dagger}$, Zhipeng Jin$^{1,3}$, Hua Tong$^{4,5}$, Shixiang Tang$^{1}$, Hongchuan Shen$^{1}$, Ning Xu$^6$, Jack Hau Yung Lo$^{1*}$, Xinliang Xu$^{2,7*}$, Lei Xu$^{1,3*}$}

\affiliation{$^1$ Department of Physics, The Chinese University of Hong Kong, Hong Kong, China
\\$^2$ The Beijing Computational Science Research Center, Beijing, China
\\$^3$ Shenzhen Research Institute, The Chinese University of Hong Kong, Shenzhen, China
\\$^4$ School of Physics and Astronomy, Shanghai Jiao Tong University, Shanghai 200240, China
\\$^5$ Department of Physics, University of Science and Technology of China, Hefei 230026, China
\\$^6$ Hefei National Laboratory for Physical Sciences at the Microscale, CAS Key Laboratory of Soft Matter Chemistry, and Department of Physics, University of Science and Technology of China, Hefei 230026, China
\\$^7$ Department of Physics, Beijing Normal University, Beijing 100875, China}

\date{\today}

\begin{abstract}
For various engineering and industrial applications, it is desirable to realize mechanical
systems with broadly adjustable elasticity in response to external environment. Here in this work, we discover a topology-correlated transition between affine and non-affine regimes in elasticity in both 2D and 3D packing-derived networks. Based on this transition, we numerically design and experimentally realize multifunctional systems with adjustable elasticity. Within one system, we achieve solid-like affine response, liquid-like non-affine response, and a continuous tunability in between. Moreover, the system also exhibits a broadly tunable Poisson's ratio from positive to negative values, which is of practical interest for energy absorption and fracture resistant materials. Our study reveals a fundamental connection between elasticity and network topology, and demonstrates its practical potential in designing mechanical systems and metamaterials.
\end{abstract}
\maketitle

Metamaterials~\cite{kadic2013} are artificial materials with unusual properties difficult to find in nature. These properties originate mainly from their structures, instead of their compositions. In particular, the mechanical metamaterials~\cite{berger2017,rock2017,reid2018,nicolaou2012,coulais2017,kane2014,zheng2014,coulais2018,goodrich2015,florijn2014} are specially designed structures with unusual elastic responses, such as negative Poisson's ratio~\cite{rock2017,reid2018}, negative compressibility~\cite{nicolaou2012}, non-reciprocity~\cite{coulais2017}, topological boundary modes~\cite{kane2014}, and ultrahigh stiffness~\cite{zheng2014}. Adjustable mechanical metamaterials whose elasticities can be self-adjusted ~\cite{coulais2018,goodrich2015,florijn2014} are of enormous interest for industrial and engineering applications. Such materials have been achieved mostly by regular combination of identical units. However, disordered jamming structures, which are physically well defined and allow rigorous mathematical analysis, were recently discovered as another promising approach~\cite{rock2017,reid2018,goodrich2015}. In this study, we realize multifunctional elastic systems with disordered jamming structures.

Jamming is a special state in material packing that occurs in a wide range of occasions, such as the jam of particles, emulsions and foams ~\cite{majmudar2007,dauchot2005,keys2007,alexander1998,silbert2002}, traffic jam, tumor progression \cite{tumor} and even in embryonic morphogenesis \cite{embryo}. At the jamming transition, rigidity arises and the system changes from fluid-like to solid-like~\cite{liu1998,olsson2007,head2009}. However, a marginally-jammed solid near jamming transition differs significantly from typical solid: for a typical solid, different moduli such as the shear modulus ($G$) and the bulk modulus ($K$) are closely related with a roughly constant ratio; while for a marginally-jammed solid the shear modulus approaches zero but the bulk modulus remains finite, making their ratio to diverge~\cite{somfai2005,makse1999,makse2000}. This $G\ll K$ behavior is similar to a liquid, which is easy to shear but difficult to compress. Extensive studies ~\cite{majmudar2007,dauchot2005,keys2007,alexander1998,silbert2002,liu1998,olsson2007,head2009,zhang2009,weitz2004,tkachenko1999,somfai2005,ellenbroek2009,wyart2005epl,wyart2005,makse1999,makse2000,ellenbroek2006,donev2005,silbert2005,zhao2011,mao2010,xu2010,liu2010,hecke2010} demonstrate that the unique elastic properties near jamming come from the non-affine response, which indicates to what extent the internal strain is independent from the external load. The non-affine properties also enable amazing designs of disordered metamaterials based on jamming \cite{rock2017,reid2018,goodrich2015}. However, an important issue remains open: can such a non-affine system turn into affine, and realize both affine and non-affine elasticities within one single system?

We first illustrate this issue in 2D. The network is derived from a 2D packing system with $N$ particles, as shown in Fig.1(a): to prevent crystallization, the particles have two sizes with the size ratio of 1:1.4 and the number ratio of 1:1. The total number of contacts between particles is $N_C$ and the contact interaction is harmonic with identical spring constant $k$. Based on this packing system, we construct a network with $N$ nodes corresponding to $N$ particle, and $N_C$ bonds (i.e., springs) corresponding to $N_C$ contacts (see SI-\RNum{1}). As illustrated in the upper panel of Fig.1(a), we can vary the contact number by numerically inflating all particles. As a result, more particles previously not in contact can touch each other, which corresponds to adding more bonds to the network, as illustrated in the lower panel. During the particle inflation process, we keep all particle or node positions fixed and identical to the jamming point configuration, which makes the experimental operations simple and practical as demonstrated later.

Such a network satisfies the modified Maxwell isostatic theorem~\cite{Calladine1978}:
\begin{equation}
dN-N_C=f(d)+N_0-N_S
\end{equation}
where $d$ is the dimension, $f(d)=d(d+1)/2$, $N_0$ is the number of floppy modes, and $N_S$ is the number of self-stress-states~\cite{wyart2008,lubensky2015}. The coordination number, $z=2N_C/N$, gives the average number of bonds per node. At the jamming point $N_0=N_S$ and we obtain the isostatic condition: $z_C=2N_C/N=2d-2f(d)/N\approx 2d$ for large enough $N$.

Above the jamming point (i.e., $z>z_C$) the system changes from fluid-like to solid-like. However, a marginally-jammed solid differs significantly from a typical solid. The underling reason is the affine versus non-affine elastic response. The affine response comes from a linear correlation between the internal displacement field and the external strain at the boundary, which typically appears in ordered or homogeneous structures. In disordered systems, however, their non-regular structures can disturb and destroy this linear correlation, and produce non-affine response. In general, ordered or homogeneous systems are typically affine while disordered or inhomogeneous systems are non-affine. However, in this study we realize both affine and non-affine elasticities within one disordered jamming system.

To realize the affine elasticity in jamming system, we minimize the non-affinity or the characteristic length $l^*$, which is the typical size of a patch cut out from the system to create a floppy mode~\cite{wyart2008,ellenbroek2009,wyart2005,wyart2005epl} (see SI-\RNum{4}). According to Eq.(1), $l^*$ reaches minimum when the system reaches complete triangulation, as shown in Fig.1(b) (see Methods). Moreover, the 2D topological invariant, Euler characteristic, further gives $z=2N_C/N=6-12/N\approx6$ at this complete triangulation (see Methods). Therefore, we identify a critical coordination number, $z_{aff}=6$, at which the entire system reaches complete triangulation and the affine property reaches maximum.

For direct visualization, we plot the internal displacement fields under external strain for $z<6$ and $z_{aff}=6$ in Fig.1(c). For $z<6$ the internal fields look rather random, while at $z_{aff}=6$ a linear correlation appears. Despite the underlying disordered structure, the system at $z_{aff}=6$ exhibits displacement fields almost identical to regular systems, confirming its affine nature (see SI-II for comparison with high-packing-density systems).

Above $z_{aff}=6$, cross or intersecting bonds start to emerge and the system is not a planar graph anymore~\cite{trudeau1993,nakamura2008,gimenez2009}: the topology changes and the Euler characteristic breaks down. The system also goes beyond the typical packing scenario, as packed particles can only touch nearby neighbors and never penetrate through them to form cross bonds. Therefore, this transition corresponds to a fundamental change in both topology and packing scenario. To reach the cross-bond regime, we first connect all nearest neighbors in the system to reach complete triangulation, and then connect the next nearest neighbors with cross bonds.

We further show that the minimum non-affinity achieved at $z_{aff}=6$ actually approaches zero and the transition is affine. To quantify non-affinity, we define a dimensionless energy difference between the ideal affine displacement field and the actual displacement field: $\Delta E=(E_{aff}-E)/E_{aff}$. Here $E_{aff}$ is the elastic energy stored in the affine displacement field mandatorily imposed according to a certain external strain, and $E$ is the energy stored in the real displacement field under the same external strain. For any non-affine system, the mandatorily-imposed affine displacement field is not a force equilibrium state and requires more energy than the real displacement field. Thus $\Delta E$ is non-negative and only approaches zero as the system approaches affine elasticity. In Fig.1(d) left panel we plot $\Delta E$ versus $\Delta z=z-z_C$ ($z_C=4$ for 2D) in our system. For both shear and compression, there is a transition at $\Delta z=2$ or $z_{aff}=6$: to its left $\Delta E$ drops rapidly towards zero, and to its right $\Delta E$ is locked around a small value about $0.1$. This transition thus enables either a significant adjustment or a lock-in capability. Moreover, $\Delta E$ approaches zero at $z_{aff}=6$, indicating a non-affine to affine transition. We show its configuration in Fig.1(d) right panel.

For comparison, we also calculate $\Delta E$ in a completely random system in Fig.1(e). Once again we observe a kink at $\Delta z=2$, which separates two distinct regimes and reveals a general feature that the topology change at $\Delta z=2$ produces two elastic regimes. However, Fig.1(e) also differs significantly from Fig.1(d): at $\Delta z=2$, $\Delta E$ is far above zero and thus the kink is not an affine transition. In fact, this random system remains highly non-affine throughout the entire $\Delta z$ range, which could be a general feature for common disordered systems.

Therefore, the special combination of particle positions at the jamming point and bond connections at $\Delta z=2$ leads to a non-affine to affine transition. To verify it, we perturb our system in Fig.1(f): as each particle is randomly displaced from the jamming configuration by larger and larger distance $d$, the non-affinity $\Delta E$ at $\Delta z=2$ rises rapidly from zero. This unambiguously shows that the affine transition correlates closely to the jamming configuration, and perturbations in general destroy this affine transition. The underlying reason is probably the high local uniformity of the jamming configuration, which is similar to ordered lattice structures but in sharp contrast to the random system (see right panels of Fig.1d-e). Thus the jamming configuration's locally-uniform node positions coupled with triangulated bond connections produce affine elasticity, and realize this ordered structure's property in a disordered system (see Fig.SI-3 in SI-III).

Apparently, besides the jamming transition at $z_C=4$, there is another non-affine to affine transition at $z_{aff}=6$. Our simulations on bulk and shear moduli $K$ and $G$ can directly verify this transition, as shown in Fig.2(a): two renormalized moduli, $K/K_{max}$ and $G/G_{max}$, are plotted as a function of $\Delta z$. Clearly near $\Delta z=0$, $G$ approaches zero while $K$ remains finite, making their ratio, $K/G$, to diverge (i.e., highly non-affine). However, for $\Delta z>2$ or $z>6$, the two curves merge together, indicating a constant ratio of $K/G=K_{max}/G_{max}$ that is the affine-like behavior. The inset further shows that only in large enough systems, $N>500$, $z_{aff}$ stabilizes at 6. Fig.2(b) plots the actual ratio, $K/G$, versus $\Delta z$ for an $N=1024$ system: for $\Delta z<2$, $K/G \sim \Delta z^{-1/2}$, which is different from the result of $K/G\sim\Delta z^{-1}$ reported in previous studies ~\cite{hecke2010,ellenbroek2009epl} due to our different bond-addition protocol; while for $\Delta z\ge 2$, $K/G\equiv2$ and the system is affine-like.

We derive a universal expression for an arbitrary modulus, $M$, in both affine and highly non-affine situations (see SI-\RNum{5}):
\begin{equation}
M\sim kz\int^{\pi}_{0}\cos^2\alpha \cdot P(\alpha)d\alpha \label{moduli estimation}
\end{equation}
with $\alpha$ the angle between an arbitrary bond and its two nodes' relative displacement due to external load (see Fig.2(c)), and $P(\alpha)$ the probability distribution of $\alpha$ in the entire system~\cite{ellenbroek2006}. We theoretically calculate $K$ and $G$ with Eq.(2) and compare with numerical simulations in Fig.2(d) and (e), and observe a good agreement. Therefore, $P(\alpha)$ in Eq.(2) fundamentally explains affine and non-affine behaviors. In the non-affine regime, $P(\alpha)$ varies differently in $K$ and $G$~\cite{ellenbroek2006,ellenbroek2009epl}, making $K/G$ varies with $z$. In the affine regime, however, $P(\alpha)$ is fixed and thus $K/G$ remains unchanged.

Combining affine and non-affine behaviors, we design networks with both types of tunability: our design is well above the jamming point $z_C=4$ and closer to $z_{aff}=6$, typically at $z>5$ to ensure enough locally-triangulated regions. One specific design is illustrated in the left panel of Fig.3(a) with $z=5.28$. Clearly this network contains two distinct local structures: the affine regions with complete triangulation and the non-affine regions close to the jamming point structure. These two types of regions exhibit distinct local properties (see Fig.SI-6 in SI), which form the basis of realizing the two-way tunability.

The non-affine regime enables us to adjust the ratio, $K/G$, across a broad range. Because some bonds make more contributions to $K$ and some are more important to $G$ \cite{goodrich2015}, deleting or adding such bonds can significantly change one modulus while keep the other one stable. We first reduce $K$ significantly by deleting the bonds important to $K$, as shown by the first two panels of Fig.3(a). The effect is shown in Fig.3(b) main panel: as the most important bonds are continuously deleted, $K$ drops dramatically while $G$ is relatively stable. It even reaches the unusual situation of $K<G$, as shown in the upper inset. Correspondingly, their ratio $K/G$ decreases significantly as shown in the lower inset. Because $K/G$ typically correlates to the Poisson's ratio, $\nu$, this also tunes $\nu$ from positive to negative, as shown in Fig.3(c) and Movie-1. Therefore, our network realizes non-affine elasticity and achieves a broad tunability in both $K/G$ and $\nu$.

We then demonstrate the affine-like behavior with a constant $K/G$ in the same network, based on the middle configuration of Fig.3(a). Clearly this configuration contains many triangulated regions with $z=6$ locally. According to our previous result, when the global structure reaches complete triangulation, the system becomes an affine solid and $K/G\equiv2$ is fixed. We now show that the elasticity is also correlated to local structures~\cite{wyart2008} (See SI-\RNum{6}). When cross bonds are added to the locally-triangulated regions, as indicated by the red bonds in Fig.3(a) right panel, both $K$ and $G$ increase while their ratio, $K/G$, remains a constant, as shown in Fig.3(d). Correspondingly, the Poisson's ratio $\nu$ remains unchanged, as shown in Fig.3(e) and Movie-2. More interestingly, now $K/G$ is not just fixed at 2 but can be locked at an arbitrary value, such as 0.5 in this example, which provides an extra dimension of tunability (See SI-\RNum{7}).

Similar to Fig.3(a), we can also reduce $G$ while keep $K$ relatively stable by deleting the bonds important to $G$ only, as shown in Fig.3(f). This procedure decreases $G$ by two orders of magnitude while keeps $K$ relatively stable, as shown in Fig.3(g). Due to increasing anisotropy, this operation does not change $\nu$ significantly (see Fig.3(h) and SI-\RNum{8}). Based on the configuration in Fig.3(f) middle panel, we again add cross bonds in the right panel and confirm the affine function, as shown in Fig.3(i) and (j).

Combining affine and non-affine features, our network realizes three functions: (1) keeping $K$ stable and significantly changing $G$, (2) keeping $G$ stable and significantly changing $K$, and (3) keeping $K/G$ stable and simultaneously changing $K$ and $G$. All functions also apply to other pairs of moduli and thus achieve a broadly-adjustable elasticity (see SI-\RNum{7}).

Next we realize this powerful system experimentally. The key issue is to eliminate the bending energy and realize pure harmonic interaction theoretically assumed. We use identical springs inside acrylic tubes as bonds and attach multiple bonds onto one smooth rod that behaves as a node. All bonds can rotate freely around each node with negligible friction, which effectively eliminates the bending energy because all bonds prefer to rotate rather than bend. To avoid conflicts between cross bonds in the same plane, we design a multi-layer system and place bonds at different heights, as shown in Fig.4(a)-(c). For simplicity, we design a small system with $N=50$ nodes, and all bonds can be reversibly added or deleted (see Movie-3).

Because the bulk modulus $K$ cannot be measured easily, we measure the Young's modulus $E$ and the shear modulus $G$ (see Methods). We first realize the non-affine property by significantly reducing $E$ while changing $G$ more gently. As shown in Fig.4(d), the colored bonds are continuously deleted with the ones more important to $E$ first, which reduces $E$ significantly while $G$ changes more gently (see Fig.4(e)). Correspondingly, $E/G$ and $\nu$ change significantly in Fig.4(f). Similarly, we can reduce $G$ significantly while keep $E$ relatively stable by deleting another set of bonds, as shown in Fig.4(g)-(i). These non-affine behaviors agree well with the simulations in Fig.3.

We then realize the affine property in the same system. We first tune $E$ and $G$ to a desirable $E/G$, as shown by the black bonds in Fig.4(j). At this configuration, we then demonstrate the affine function by adding the cross bonds in triangulated regions, as shown by the colored bonds. This varies both $E$ and $G$ simultaneously (see Fig.4(k)), while keeps $E/G$ and $\nu$ at a constant (see Fig.4(l)). Thus we realize an independent tuning on $E$, $G$, $E/G$, and $\nu$, and achieve affine and non-affine elasticity experimentally.

Furthermore, all our 2D results can be extended to 3D. Similar to complete triangulation in 2D, when the network is fully tetrahedralized in 3D, a non-affine to affine transition appears (see Methods). Our simulation confirms this transition at $z_{aff}=12.8$ in Fig.5(b) ($z_{aff}\neq12$ due to polydisperse particle size).

Above $z_{aff}$, cross bonds start to appear. Note that in 3D they go across a face while in 2D they go across another bond, as shown in Fig.5(a). Similar to 2D, the 3D cross bonds also break a fundamental topology, the Euler-Poincar\'{e} characteristic~\cite{nakamura2008}, which is the 3D generalization of the Euler's characteristic in 2D (see SI-X). Therefore, in both 2D and 3D packing-derived networks, a fundamental link between the non-affine to affine transition and the change in topology exists.

Analogous to 2D, adding cross bonds inside locally-tetrahedralized regions can again lock the non-affinity. As shown in Fig.5(c), the non-affinity $\Delta E$ drops rapidly before $z_{aff}$ and stabilizes around a small value after $z_{aff}$, confirming the non-affine and affine regimes. Moreover, we can tune $\Delta E$, $K/G$ or $\nu$ to a desirable value, and then lock it by adding cross bonds, as shown by the triangle symbols and the inset. Therefore, all the affine and non-affine functions in 2D are realized in 3D.

For practical applications, realizing actual 3D metamaterials is essential. Using 3D printing, we achieve such metamaterials in Fig.5 (d)-(f): the $5\times5\times5$ node positions of these three networks are identical and from the same packing configuration, differing only in their bond number or $z$. From (d) to (e), $z$ increases from 7.696 to 9.312 by adding non-cross bonds, which tunes $K/G$ and $\nu$. From (e) to (f), however, $z$ increases from 9.312 to 10.432 by adding cross bonds, which locks $K/G$ and $\nu$ while at the same time strengthening the system by increasing both $K$ and $G$.

To verify their performance, we apply identical external strain to these three networks, and compare their internal strain fields which essentially determine the elastic responses such as non-affinity and Poisson's ratio. As illustrated in Fig.5(g)-(i), row 1 and 2 exhibit significant difference in internal strain, demonstrating distinct elastic responses due to our tuning operation; however, row 2 and 3 show almost identical internal strain, verifying the same elastic responses due to our locking operation (also see Movie-4 and Movie-5). The forces generated by this strain are also measured as 7.86 N, 10.47 N, and 12.24 N respectively, confirming the expected increase of system rigidity with respect to $z$. Note that our 3D-printed bonds do exhibit bending force, which does not match the theoretical assumption perfectly. However, Fig.5(g)-(i) still show a nice experimental outcome, demonstrating the robustness of our design, which paves the way for realizing actual metamaterials beyond ideal spring networks. We further note that all bonds can be 3D-printed as detachable to realize all functions in a single system (see Movie-6). In addition, the size range of the bonds and the overall system may vary multiple orders to satisfy various practical requirements. The plasticity and failure behaviors can also withstand a broad range of external stress due to its heterogeneous structure, and avoid the sudden failure of ordered lattice systems, which are composed mostly by identical bonds that may fail simultaneously.

To conclude, we discover a non-affine to affine transition at the point of topology change. Based on this fundamental transition, we realize networks broadly tunable in $K$, $G$, $K/G$ and $\nu$, and achieve both affine and non-affine elasticities within one system. Moreover, such systems might be self-assembled by packing granular materials and control connections either chemically \cite{li2021} or electrically with electromagnetic devices, which may greatly improve their practical applications. Our study reveals a fundamental connection between elasticity and topology, and provides a practical design principle for multifunctional mechanical metamaterials.

\noindent\textbf{Data availability.} All the raw data of the figures presented in this manuscript can be obtained by accessing to the Open Science Framework (DOI 10.17605/OSF.IO/7EQ5Z) or directly visiting https://osf.io/7eq5z/ \cite{rawdata}.

\noindent\textbf{Methods}\\
\noindent\textbf{Deriving the transition in 2D}. The non-affinity will be minimized when $l^*$ reaches minimum, i.e., no floppy mode can be created even in an arbitrary smallest patch. A smallest patch is formed by 3 nearby and non-collinear nodes and we apply the theorem of Eq.(1) onto this 3-node patch: no floppy mode means $N_0=dN-N_C-f(d)+N_S=0$. Plugging in $d=2, N=3, f(d)=3, N_S=0$ (no self-stress-state can exist in 3 non-collinear nodes), we get $N_C=3$. This means 3 bonds are required for this 3-node patch, i.e., a triangle is needed for zero floppy mode, as shown in Fig.1(b) left panel. If we eliminate one bond as shown in the right panel, the 3-node patch can now freely rotate without costing any energy, which creates one floppy mode. Therefore, to minimize $l^*$ and non-affinity, 3 bonds are required for every 3-node patch and the system reaches the complete triangulation.

A completely triangulated configuration contains no intersecting or cross bonds and belongs to the planar graph, which obeys the topological invariant - the Euler characteristic: $N_f-N_C+N=2$~\cite{trudeau1993,nakamura2008}, with $N_f$ the number of faces enclosed by edges or bonds. For a triangulated large-$N$ system whose boundary influence can be neglected, each face is surrounded by 3 bonds, and each bond is shared by 2 faces, and thus $3N_f=2N_C$. Plugging $N_f=2N_C/3$ into the Euler characteristic, we get: $N_C=3N-6$ and thus $z=2N_C/N=6-12/N\approx6$. Thus at $z_{aff}=6$ the system reaches the non-affine to affine transition.\\

\noindent\textbf{Deriving the transition in 3D.} The smallest patch in 3D is formed by 4 non-coplanar nodes and we apply the Maxwell isostatic theorem, $N_0=dN-N_C-f(d)=0$, to it. Plugging in $d=3$, $N=4$, $N_S=0$, and $f(d)=6$, we get $N_C=6$. Thus the smallest patch without floppy mode is a tetrahedron with four nodes and six bonds. Thus when the network is fully tetrahedralized, a non-affine to affine transition will appear.\\

\noindent\textbf{Measuring $E$ and $G$ in the spring system.} We first measure the stress by fixing one boundary of the system and moving the opposite boundary with $5\%$ strain, in either parallel ($G$) or perpendicular ($E$) direction. We then divide the stress with the strain and obtain the modulus. Due to boundary effect, these measurements still have about $10\%$ deviations from the rigorous definition (see SI-\RNum{9}) but are good enough experimental approximations.
\\

\noindent\textbf{Data availability} All the raw data of the figures presented in this manuscript can be obtained by accessing to the Open Science Framework (DOI 10.17605/OSF.IO/7EQ5Z) or directly visiting https://osf.io/7eq5z/.
\\

\noindent\textbf{Code availability} All custom computer code or algorithm used to generate results that are reported in the paper are available upon request.
\\

\noindent\textbf{Acknowledgements} The experiments are performed in the Chinese University of Hong Kong, and we acknowledge the computational support from the Beijing Computational Science Research Center. L. X. acknowledges the financial support from NSFC-12074325, Guangdong Basic and Applied Basic Research Fund 2019A1515011171, GRF-14306518, CRF-C6016-20G, CRF-C1018-17G, CUHK United College Lee Hysan Foundation Research Grant and Endowment Fund Research Grant, CUHK direct grant 4053354, X. X. acknowledges the financial support from NSFC 11974038 and U1930402, X. S. acknowledges the financial support from Guangdong Basic and Applied Basic Research Foundation 2019A1515110211, and Project funded by China Postdoctoral Science Foundation 2020M672824.

\noindent\textbf{Author contributions} X. S. and C. F. contribute equally to this research, L. X. conceived the research, X. S., C. F., J. H. Y. L., X. X. and L. X. designed the research, X. S. performed most theoretical and numerical analysis, C. F. performed most experiment, Z. J., S. T., H. S., H. T., and N. X. helped in the experiment or the simulation, X. S., C. F. and L. X. prepared the manuscript, L. X. and X. X. supervised the research.

\noindent\textbf{Competing interests} The authors declare no competing financial interests.

\noindent\textbf{Correspondence and requests for materials} should be addressed to L. X. (xuleixu@cuhk.edu.hk), X. X. (xinliang@csrc.ac.cn), and J. H. Y. L. (hylo@cuhk.edu.hk).

\clearpage
\newpage

\begin{figure*}[t]
\begin{center}
\centerline{\includegraphics[width=0.99\linewidth]{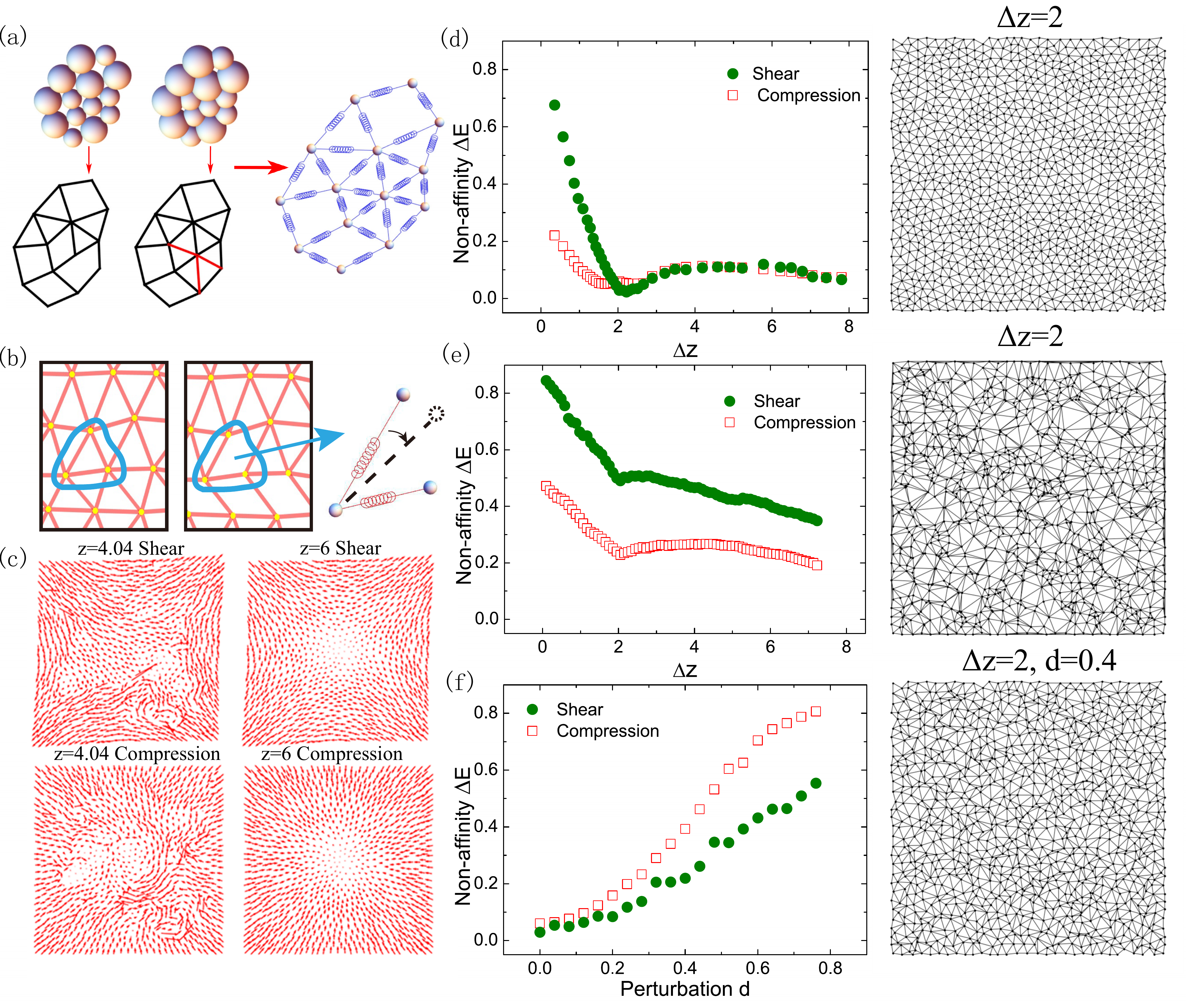}}
\caption{\textbf{Non-affine to affine transition in packing derived networks.} \textbf{a}, converting a particle packing configuration into a spring network. Particles correspond to nodes and contacts correspond to bonds. We can vary the packing density by inflating particles (upper panels), which increases the bonds of the network (lower panels). \textbf{b}, for a smallest patch with 3 nodes cut out of the system, 3 bonds are required to eliminate the floppy mode. If one bond is missing, one floppy mode by free rotation will be created. \textbf{c}, displacement field before ($z=4.04$) and at ($z=6$) the non-affine to affine transition under shear and compression stress. The field is random before the transition and becomes linear at the transition. \textbf{d}, the non-affinity $\Delta E$ approaches zero at $\Delta z=z-z_C=2$, indicating an almost ideal affine transition in our packing derived network, whose configuration is shown in the right panel. \textbf{e}, $\Delta E$ never approaches zero in a random network, indicating its non-affine nature. Similar to \textbf{d}, a universal kink appears at $\Delta z=2$ which separates two distinct elastic regimes. The right panel shows the random network configuration. \textbf{f}, as every node position is perturbed by larger and larger displacement $d$, the non-affinity $\Delta E$ at the kink $\Delta z=2$ rises sharply, confirming that the affine transition uniquely occurs at the jamming point configuration, $d=0$. The right panel shows the configuration at $d=0.4$ ($d$ has the unit of small particle's diameter that is close to an average bond length).}
\end{center}
\end{figure*}

\clearpage
\begin{figure*}[t]
\begin{center}
\centerline{\includegraphics[width=0.96\linewidth]{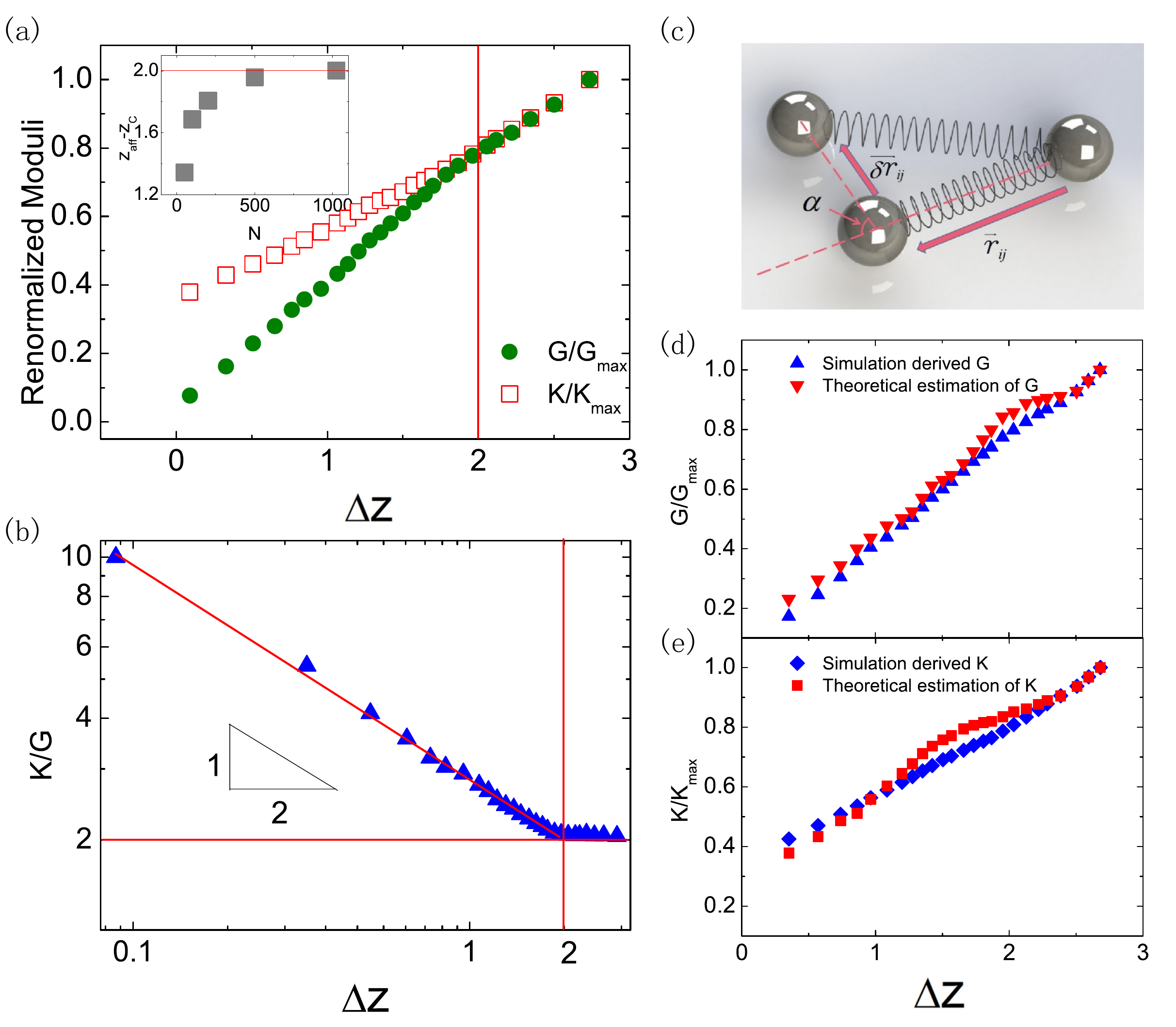}}
\caption{\textbf{Understanding the variation of moduli at single particle level.} \textbf{a}, the non-affine to affine transition at $\Delta z=2$ in an $N=1024$ system. When $\Delta z<2$, $G/G_{max}$ and $K/K_{max}$ vary differently; however they collapse onto one curve for $\Delta z > 2$. The inset shows that the transition stabilizes at $\Delta z=2$ for $N>500$ systems. \textbf{b}, $K/G\sim \Delta z^{-1/2}$ before the transition while $K/G=2$ after the transition. \textbf{c}, the schematics of the angle $\alpha$ between an arbitrary bond, $\vec{r}_{ij}=\vec{r}_{j}-\vec{r}_{i}$, and its two nodes' relative displacement under external stress, $\vec{\delta r}_{ij}=\vec{\delta r}_{j}-\vec{\delta r}_{i}$. \textbf{d}, comparison between theory and simulation for renormalized shear modulus, $G/G_{max}$. \textbf{e}, comparison between theory and simulation for renormalized bulk modulus, $K/K_{max}$. In both \textbf{d} and \textbf{e}, the agreement is reasonable with the largest deviation around $10\%$ to $15\%$.}
\end{center}
\end{figure*}

\clearpage
\begin{figure*}[t]
\begin{center}
\centerline{\includegraphics[width=0.92\linewidth]{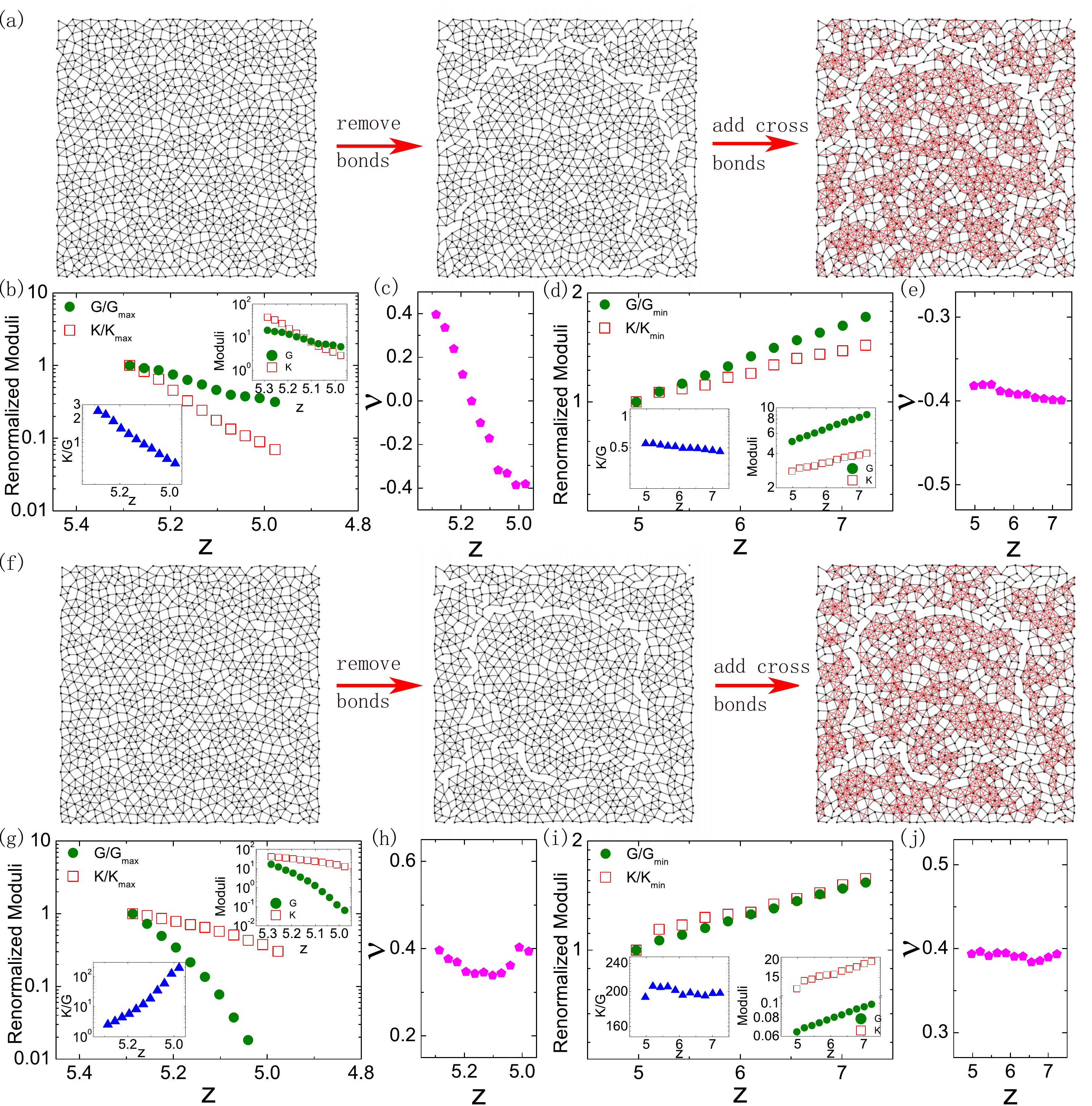}}
\caption{\textbf{The numerical design of a network with both affine and non-affine tunability.} \textbf{a}, panel 1: the original network with $z=5.28$, panel 2: achieving non-affine tunability by removing the bonds important to $K$, panel 3: realizing affine tunability by adding the red cross bonds. \textbf{b}, with the non-affine operation of bond removal, $K$ reduces significantly while $G$ changes more gently (note that $z$ axis decreases with this operation). Interestingly, $K/G$ can decrease even below unity to reach the unusual situation of $K<G$ (see the two insets). \textbf{c}, the Poisson's ratio $\nu$ is also broadly tunable from positive to negative by this operation. \textbf{d}, with the affine operation of cross bond addition, both $K$ and $G$ increase but $K/G$ remains roughly a constant (note that $z$ axis increases with this operation). \textbf{e}, the Poisson's ratio $\nu$ also stays constant under this affine operation. \textbf{f}, operations similar to \textbf{a} except that the bonds important to $G$ are now removed and then the cross bonds are added. \textbf{g}, when the bonds important to $G$ are removed, $G$ reduces by two orders while $K$ remains relatively stable. Correspondingly, $K/G$ increases significantly. \textbf{h}, $\nu$ does not change much with bond removal. \textbf{i} and \textbf{j}, the affine operation of cross bond addition changes $K$ and $G$ simultaneously, while $K/G$ and $\nu$ remain unchanged as expected. }
\end{center}
\end{figure*}

\clearpage
\begin{figure*}[t]
\begin{center}
\centerline{\includegraphics[width=0.87\linewidth]{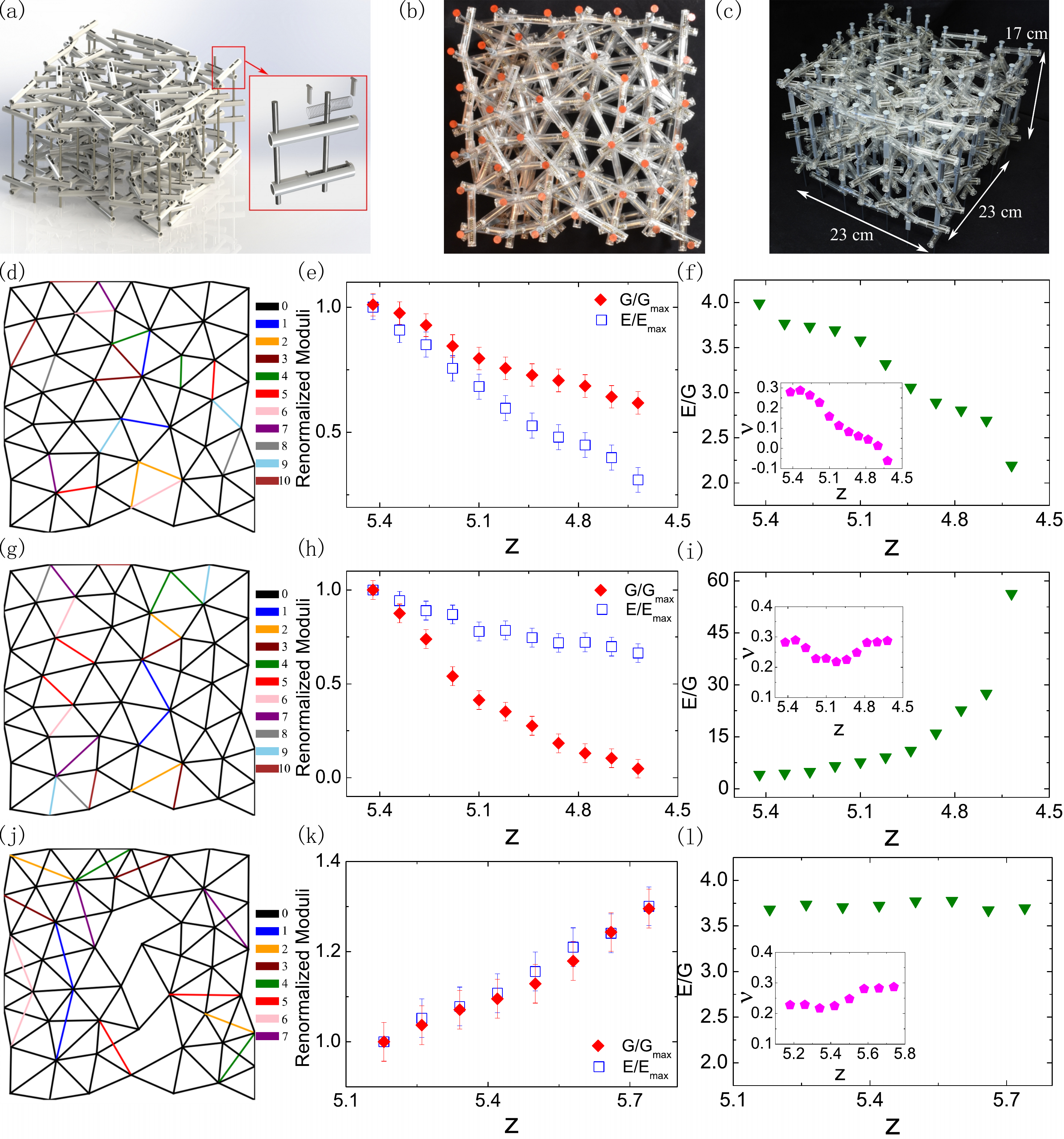}}
\caption{\textbf{Experimental realization of affine and non-affine tunability.} \textbf{a}, schematics of our experimental system. The bonds are identical springs confined in acrylic tubes. The nodes are long rods for the bonds to hold on. The system is designed as multi-layered to avoid conflicts between crossing bonds, and vertically symmetric to eliminate net torques. \textbf{b}, top view image of the actual system with all nodes ($N=50$) labeled in red. $z=5.42$ in this system. \textbf{c}, side view image of the actual system. \textbf{d}, realizing the non-affine property by removing the colored bonds which are critical to $E$ in the order from 0 to 10, with more important ones removed earlier. \textbf{e}, the removal of bonds decreases $E$ significantly but much less in $G$. Note that both $E$ and $G$ are measured over two perpendicular directions and then averaged. \textbf{f}, the bond removal achieves significant adjustment on $E/G$ and $\nu$. \textbf{g}, similar operation of bond removal for the ones critical to $G$ only (shown in the order from 0 to 10). \textbf{h}, $G$ reduces significantly while $E$ decreases gently. \textbf{i}, the bond removal increases $E/G$ significantly but $\nu$ does not change much. \textbf{j}, realizing the affine property by adding the colored cross bonds in the order from 0 to 10. \textbf{k}, with bond addition both $E$ and $G$ increase at the same rate. \textbf{l}, $E/G$ and $\nu$ stay unchanged, as expected for an affine system.}
\end{center}
\end{figure*}

\clearpage
\begin{figure*}[t]
\begin{center}
\centerline{\includegraphics[width=0.96\linewidth]{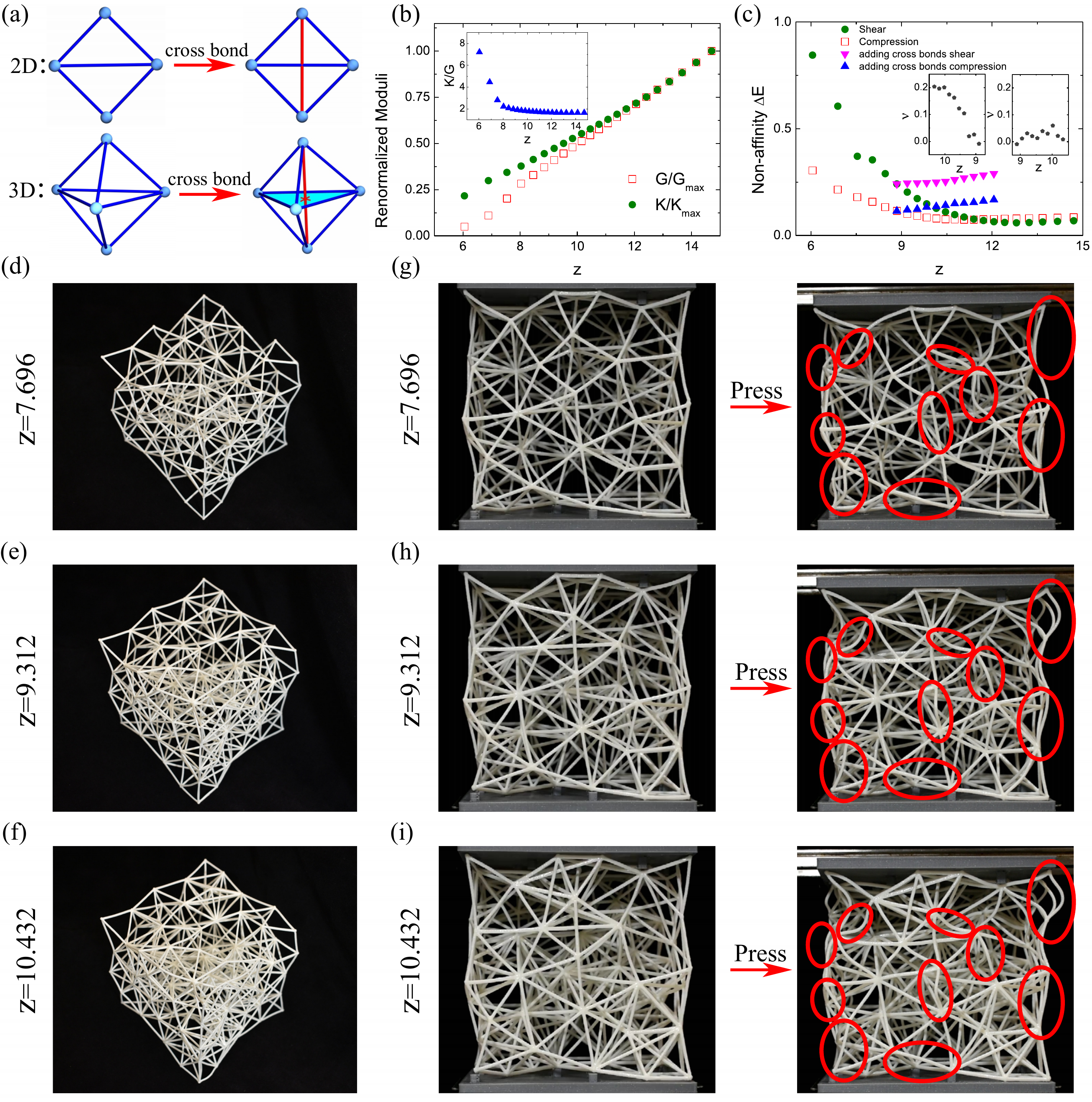}}
\caption{\textbf{Extending the 2D results into 3D.} \textbf{a}, schematics showing the cross bond (denoted in red) in 2D and 3D respectively. In 2D it goes across a bond and in 3D it goes across a face. \textbf{b}, $G/G_{max}$ and $K/K_{max}$ merge together above $z_{aff}=12.8$ in a $32\times32\times32$ nodes packing derived network, confirming the affine transition in 3D. Inset shows their ratio $K/G$. \textbf{c}, the non-affinity $\Delta E$ under both shear and compression strain decreases with $z$ and stabilizes in the affine regime of $z\ge12.8$ (circles and squares). The triangles show the locking of non-affinity by adding cross bonds. Inset shows the tuning and locking of Poisson's ratio $\nu$. \textbf{d-f}, side view of the 3D printed $5\times5\times5$ nodes networks at $z=7.696$, $z=9.312$, and $z=10.432$ respectively. They have identical node positions but different bond numbers or z. From the top to the middle panel, non-cross bonds are added which tunes the affinity $K/G$ and the Poisson's ratio $\nu$. From the middle to the bottom panel, cross bonds are added which locks $K/G$ and $\nu$. \textbf{g-i} compare the internal strain fields of the three networks under an identical strain: \textbf{g} and \textbf{h} exhibit different internal strain due to the tuning operation, while \textbf{h} and \textbf{i} exhibit similar internal strain due to the locking operation.}
\end{center}
\end{figure*}

\end{document}